\documentclass[
  aps,physrev,reprint,superscriptaddress,
  amsmath,amssymb
]{revtex4-2}

\usepackage{graphicx}
\usepackage{bm}
\usepackage{ulem}
\usepackage{braket} 
\usepackage{tikz}
\usepackage[hidelinks]{hyperref}
\graphicspath{{figures/}}

\DeclareRobustCommand{\hexagon}{\mathord{\text{%
  \tikz[x=1.15ex,y=1.15ex,line width=0.07ex,baseline=0pt]{%
    \draw (0,0.5) -- (0.289,1) -- (0.866,1) -- (1.155,0.5)
      -- (0.866,0) -- (0.289,0) -- cycle;}}}}

\begin{document}  

\title{Magnetism on emergent molecular-orbital lattices}
\author{Zhan Liu}
\thanks{These authors contributed equally to this work.}
\affiliation{Beijing National Laboratory for Condensed Matter Physics and Institute of Physics, Chinese Academy of Sciences, Beijing 100190, China}
\affiliation{University of Chinese Academy of Sciences, Beijing 100049, China}

\author{Chenrui Wang}
\thanks{These authors contributed equally to this work.}
\affiliation{Beijing National Laboratory for Condensed Matter Physics and Institute of Physics, Chinese Academy of Sciences, Beijing 100190, China}
\affiliation{University of Chinese Academy of Sciences, Beijing 100049, China}

\author{Chen Fang}
\email[Contact author: ]{cfang@iphy.ac.cn}
\affiliation{Beijing National Laboratory for Condensed Matter Physics and Institute of Physics, Chinese Academy of Sciences, Beijing 100190, China}

\begin{abstract}
In solids, strong hybridization between nearby atomic orbitals forms molecular orbitals (MOs), the range of which covers a cluster of atoms. The centers of molecular orbitals form an emergent lattice generally different from the atomic lattice. For partially filled MOs, electron interaction may induce a new type of magnetism on this emergent lattice where the local moments come from the molecular orbitals instead of single atoms. In this work, we show in a ruby-lattice tight-binding Hubbard model that frustrated and unfrustrated antiferromagnetism may appear on the same atomic lattice and at the same filling, as the emergent molecular-orbital lattice changes from triangular to honeycomb upon changing the hopping parameters. Moreover, we show that general molecular orbital magnetism may be experimentally verified in the low-energy dynamic structure factor: the molecular form factors, as the Fourier transforms of the molecular spin-density, modulate the low-energy dynamic structure factor across different Brillouin zones. Zeros of these form factors suppress low energy spectral weight irrespective of the magnetic order formed by the molecular moments.
This is exemplified by random phase approximation calculations for the ruby-lattice model: the structure factor integrated over low energies exhibits minima around molecular form-factor zeros, while its ratios between selected equivalent momenta in different Brillouin zones agree with ratios of squared form factors.
\end{abstract}

\maketitle

\section{Introduction}
\label{sec:introduction}
Magnetism in solids can be described in terms of local moments or itinerant electrons. In the conventional local-moment picture, magnetic moments are associated with individual ions~\cite{Anderson1959Superexchange}. 
Strong hybridization between neighboring atoms, however, can spread the electronic wave functions over several atomic sites, giving rise to molecular orbitals (MOs). When these molecular orbitals are partially filled, interactions may induce local moments carried by extended MOs rather than individual ions~\cite{Walters2009Covalency,Mazin2012MolecularOrbitalCrystal,Foyevtsova2013MagneticInteractions,Kim2014MolecularJeff,Kim2020MolecularMott,Petersen2023ResonatingHoles}.
Experimental evidence for such molecular magnetic units has emerged in both insulating and metallic systems~\cite{Sheckelton2012ValenceBond,Jeong2017MolecularJeff,Sheckelton2017ClusterSinglet,Pasco2019TunableMagneticTransition,Yang2022BondOrdering,Gao2023SingleBandMott,Jin2023MagneticMolecularOrbitals,Grahlow2024ClusterMott}.
In the cluster Mott insulator GaTa$_4$Se$_8$, resonant inelastic x-ray scattering identifies quasimolecular moments distributed over Ta$_4$ tetrahedra~\cite{Jeong2017MolecularJeff,Magnaterra2024QuasimolecularMoments}.
In MnSi, neutron scattering combined with Wannier analysis reveals magnetic molecular orbitals extending over three neighboring Mn atoms~\cite{Jin2023MagneticMolecularOrbitals}.
The spatial spin-density profiles of molecular moments leave observable signatures in momentum dependence of scattering intensities through their form factors~\cite{Mourigal2014MolecularMagnetism,Iida2019TriangularMagnetism}. Such signatures have been reported for Mo$_3$ clusters in Li$_2$InMo$_3$O$_8$~\cite{Iida2019TriangularMagnetism}, V$_4$ tetrahedra in GaV$_4$S$_8$~\cite{Pokharel2021SpinDynamics}, and Cu--ligand units in covalent cuprates~\cite{Mazurenko2015CovalencyFormFactor}.
These observations establish spatially extended molecular units as physically relevant building blocks of magnetism in solids~\cite{Dally2020SpinDensity,Revelli2022QuasimolecularStructure,Mazurenko2015CovalencyFormFactor}.

The centers of the MOs carrying these moments define an emergent lattice, whose geometry, together with interactions between molecular moments, can support rich magnetic behavior~\cite{Merino2018SpinMolecularOrbit}, including long-range magnetic orders~\cite{Haraguchi2015ClusterMagnetism,Iida2019TriangularMagnetism,Nikolaev2021ClusterMott,Kezsmarki2015Skyrmion,Ruff2015Multiferroicity,Dally2020SpinDensity} as well as short-range correlations~\cite{Iida2019TriangularMagnetism,Nikolaev2021ClusterMott}. Representative studies of molecular-orbital magnetism have explored Haldane phases~\cite{Janani2014EffectiveTheories}, spin-1 and spin-0 cluster Mott insulators~\cite{Nourse2021SpinMottTransition} and quantum spin liquids~\cite{Law2017TaS2SpinLiquid,He2018SpinonFermiSurface,AkbariSharbaf2018TunableSpinLiquidity,Chen2022MolecularMagnets}.
The formation of different MOs can also change the emergent lattices and hence magnetic phases on them. Ref.~\cite{Nourse2021MultipleInsulatingStates} investigated rich phases on the decorated honeycomb lattice, where different MOs yield a possible quantum spin liquid on a frustrated kagome lattice and N\'eel order on an unfrustrated honeycomb lattice. However, these qualitatively distinct states occur at different fillings. A natural question is therefore whether at fixed filling, changing MOs alone can switch between frustrated and unfrustrated antiferromagnetism on the same atomic lattice. Experimental verification of such MO magnetism is equally important, yet obtaining an MO interpretation directly from experimental spectra remains challenging. Existing evidence usually relies on detailed analysis of internal molecular excitations or interpreting measured spectra with electronic-structure calculations and microscopic models~\cite{Mourigal2014MolecularMagnetism,Jeong2017MolecularJeff,Pokharel2021SpinDynamics,Magnaterra2024QuasimolecularMoments}. This raises a complementary question: are there general signatures in experimental spectra that can provide direct clues to MO magnetism?
 
\begin{figure}[tb]
   \centering
   \includegraphics[width=\columnwidth]{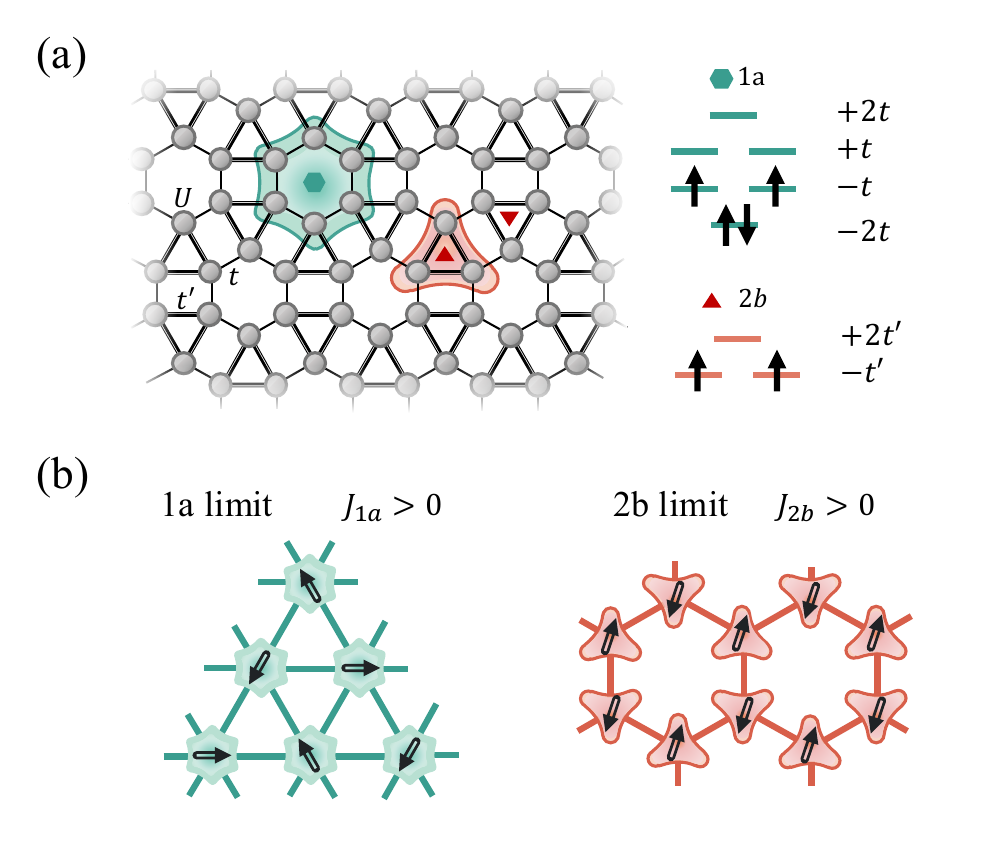}
   \caption{Emergent frustrated and unfrustrated molecular-orbital magnetism in the ruby-lattice Hubbard model. The filling is fixed at four electrons per unit cell $(n_e=4)$. (a) Ruby lattice and MO level schemes of two kinds of clusters. Green hexagonal and red triangular shaded regions indicate the corresponding hexamer and trimer molecular units centered at 1a and 2b Wyckoff positions respectively. A hexamer contains four electrons and a trimer two; the arrows indicate a representative spin-1 configuration in each cluster. (b) Emergent spin-1 AFM Heisenberg models on the frustrated triangular lattice of hexamers (left) and the unfrustrated honeycomb lattice of trimers (right), yielding $120^\circ$ order and collinear N\'eel order respectively.}
   \label{fig:ruby schemes}
 \end{figure} 

Here, we demonstrate MO magnetism in a ruby-lattice Hubbard model [Fig.\ref{fig:ruby schemes}(a)], where changing molecular units enables a switch between frustrated and unfrustrated magnetism at fixed filling, and establish two spectroscopic fingerprints for identifying MO magnetism in general. In the ruby-lattice model, varying the hopping amplitudes forms MOs on hexamers or trimers, with interactions stabilizing a spin-$1$ molecular moment distributed over each cluster. Their centers define a frustrated triangular or unfrustrated honeycomb lattice, as shown in Fig.\ref{fig:ruby schemes}(b). In the limits of weak intercluster coupling, we derive antiferromagnetic (AFM) exchange interactions that favor $120^\circ$ and N\'eel order on these lattices, respectively. Mean-field calculations further indicate that both orders persist beyond these limits. More generally, the spatial profiles of molecular moments are encoded in their form factors $F(\boldsymbol{q})$, the Fourier transform of molecular spin density. For $\omega\ll \Delta$, where $\Delta$, the molecular spin stiffness, is the intracluster gap between the ground and first excited states, we show that the dynamic structure factor $\mathcal{S}(\boldsymbol{q},\omega)$ is modulated by these form factors, yielding two spectroscopic fingerprints observable in inelastic neutron scattering. The first is the strong suppression of dynamic structure factor at common zeros of form factors throughout this low-energy range. Second, for suitable momenta $\boldsymbol{q}$ and reciprocal lattice vector $\boldsymbol{G}$,
\begin{equation}
\dfrac{\mathcal{S}(\boldsymbol{q}+\boldsymbol{G},\omega)}{\mathcal{S}(\boldsymbol{q},\omega)}\simeq \dfrac{|F(\boldsymbol{q}+\boldsymbol{G})|^2}{|F(\boldsymbol{q})|^2}, 
\end{equation}
yielding frequency-independent spectral ratios within this window, even though the individual spectra vary with frequency. These signatures are insensitive to detailed spin correlations and directly reflect features of molecular spin density, providing experimental clues to MO magnetism. Candidate MO descriptions can then be tested against these signatures using cluster ground states alone, without calculating collective spin dynamics or intracluster excitations. We demonstrate this approach in the ruby-lattice Hubbard model, where dynamic structure factors calculated within random phase approximation (RPA) exhibit the predicted fingerprints in both the $120^\circ$ and N\'eel phases. 
      
The paper is organized as follows.
In Sec.~\ref{sec:warmup}, as a warm-up, a dimerized Hubbard chain illustrates how magnetic moments can reside at a bond center rather than on an atomic site.
Section~\ref{sec:ruby} studies the ruby-lattice Hubbard model and its emergent magnetic phases.
In Sec.~\ref{sec:formfactor}, we develop a low-energy description of the dynamic structure factor and examine scattering signatures of molecular moments.
Finally, Sec.~\ref{sec:discussion} summarizes our main results and discusses the promising future developments of this new type of magnetism.

\section{Warm-up}
\label{sec:warmup}
\begin{figure}[tb]
   \centering
   \includegraphics[width=\columnwidth]{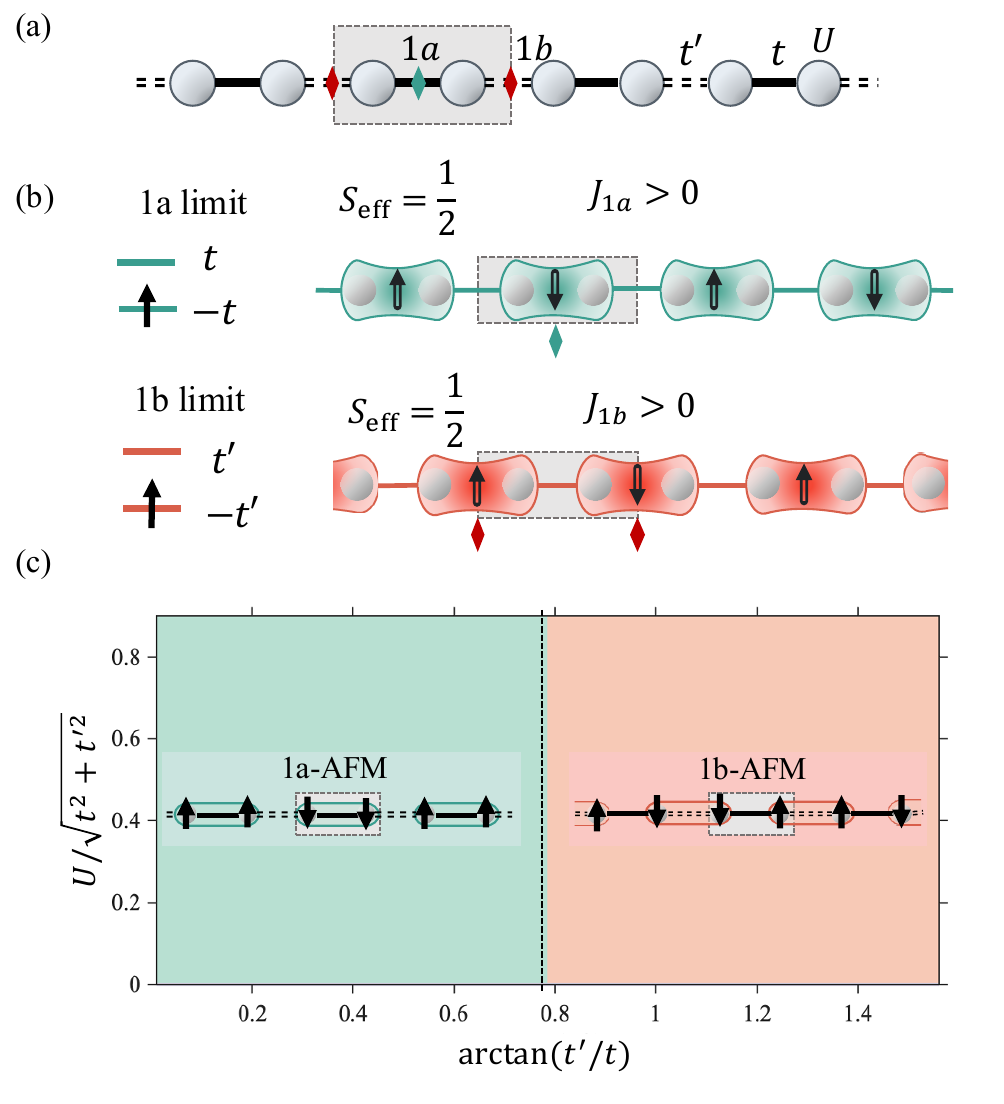}
   \caption{Molecular-orbital magnetism in a dimerized Hubbard chain with one electron per dimer. (a) Schematic of the model with alternating hopping $t$, $t'$ and onsite repulsion $U$. The dashed box indicates a unit cell, and green and red diamonds mark the $1a$ and $1b$ bond centers, respectively.
    (b) MO level schemes and effective spin-$1/2$ AFM Heisenberg chains in the $1a$ limit, $t\gg U\gg t'$ (top), and $1b$ limit, $t'\gg U\gg t$ (bottom). Each molecular moment extends over a shaded dimer centered at $1a$ or $1b$.
    (c) Mean field phase diagram. The green 1a-AFM and red 1b-AFM phases are separated by dashed line at $t'/t=1$ ($\arctan(t'/t)=\pi/4$). The insets show the corresponding atomic-site moment patterns. }
   \label{fig:1Dmodel}
 \end{figure}
As a warm-up, we illustrate molecular-orbital magnetism in a dimerized Hubbard chain, where magnetic moments are carried by bond-centered MOs. Each unit cell contains two sites denoted by $\alpha=a,b$. The intra(inter)cell hopping amplitude is $t(t')$ and the onsite repulsive Hubbard interaction is $U$.  
 The Hamiltonian for this chain is written as 
\begin{equation}
\begin{aligned}
\hat{H}_{1D}=&t\sum_{j,\sigma} (\hat{c}_{aj\sigma}^{\dagger}\hat{c}_{bj\sigma}+\text{h.c.})+t'\sum_{j,\sigma}(\hat{c}_{bj\sigma}^{\dagger}\hat{c}_{aj+1\sigma}+\text{h.c.})\\
&+U\sum_{j}(\hat{n}_{aj\uparrow}\hat{n}_{aj\downarrow}+\hat{n}_{bj\uparrow}\hat{n}_{bj\downarrow}),
\end{aligned}
\end{equation}
where $\hat{c}_{\alpha j\sigma}$ annihilates an electron with spin $\sigma$ on site $\alpha$ in unit cell $j$.  We fill this chain with one electron per unit cell ($n_e=1$). 
\par The intracell and intercell bond centers are the maximal Wyckoff positions $1a$ and $1b$ respectively [Fig.\ref{fig:1Dmodel}(a)], corresponding to two dimer limits of this chain in Fig.\ref{fig:1Dmodel}(b). When $t\gg U\gg t'$, atomic orbitals $\ket{aj}$ and $\ket{bj}$ within each unit cell hybridize into bonding and antibonding MOs centered at 1a: 
\begin{equation}
\begin{aligned}
\ket{\pm j}_{1a}=\dfrac{1}{\sqrt{2}}(\ket{aj}\pm\ket{bj}).
\end{aligned}
\end{equation}
For $t>0$, the lower bonding orbital is $\ket{-j}_{\text{1a}}$ with energy $-t$. At $t'=0$ each dimer contains one electron in the bonding MO with its spin density shared equally between the two sites. The antibonding MO lies at energy $+t$, separated from bonding orbital by a gap of $2t$. For $t\gg U$, this gap protects bonding MOs from mixing with antibonding MOs, allowing us to project onto the bonding MOs. Turning on the intercell hopping $t'$ then yields a half-filled Hubbard model on an emergent lattice of 1a dimer centers. At $U\gg t'$, interdimer charge fluctuations are suppressed, leaving an effective spin-$1/2$ moment on each dimer, whose low-energy description is the following antiferromagnetic Heisenberg model~\cite{Anderson1959Superexchange}, as the 1a limit illustrated in Fig.\ref{fig:1Dmodel}(b):
\begin{equation}
\begin{aligned}  
\hat{H}_{1a} = J_{1a}\sum_{j}\hat{\boldsymbol{S}}_j\cdot\hat{\boldsymbol{S}}_{j+1},\ J_{1a}=\dfrac{2t'^2}{U}\textgreater 0.
\end{aligned}
\label{eq: 1d 1a Hamiltonian}
\end{equation}
Its exact ground state exhibits algebraically decaying antiferromagnetic correlations~\cite{Affleck1998CorrelationAmplitude}. An analogous analysis is applied for $t'\gg U\gg t$, where atomic orbitals $\ket{bj}$ and $\ket{aj+1}$ hybridize into MOs centered at 1b. These half-filled bonding MOs then yield an AFM Heisenberg chain on a lattice of $1b$ bond centers with exchange coupling $J_{1b}=2t^2/U\textgreater 0$. Additional perturbation details in these limits are provided in the Supplemental Material~\cite{SupplementalMaterial}.
\par The two dimer limits thus demonstrate antiferromagnetically correlated molecular spins centered at 1a or 1b. The real space alignment of these moments at finite interdimer hopping is illustrated by mean-field solutions in Fig.~\ref{fig:1Dmodel}(c). The 1a-AFM solution has $m_{aj}=m_{bj}=(-1)^jm_0$, whereas the 1b-AFM solution has $m_{bj}=m_{aj+1}=(-1)^{j+1}m_0$. In both cases, the two sites within the corresponding dimer carry equal, parallel moments, reproducing the spin-density profile of the bonding MO, while neighboring dimer moments are antiparallel. The symmetry under $t\leftrightarrow t'$ and the boundary at $t'/t=1$ reflect the duality between the 1a and 1b emergent lattices. Further technical details of mean-field calculations are provided in the Supplemental Material\cite{SupplementalMaterial}. 
\begin{figure*}[t]
  \centering
   \includegraphics[width=\textwidth]{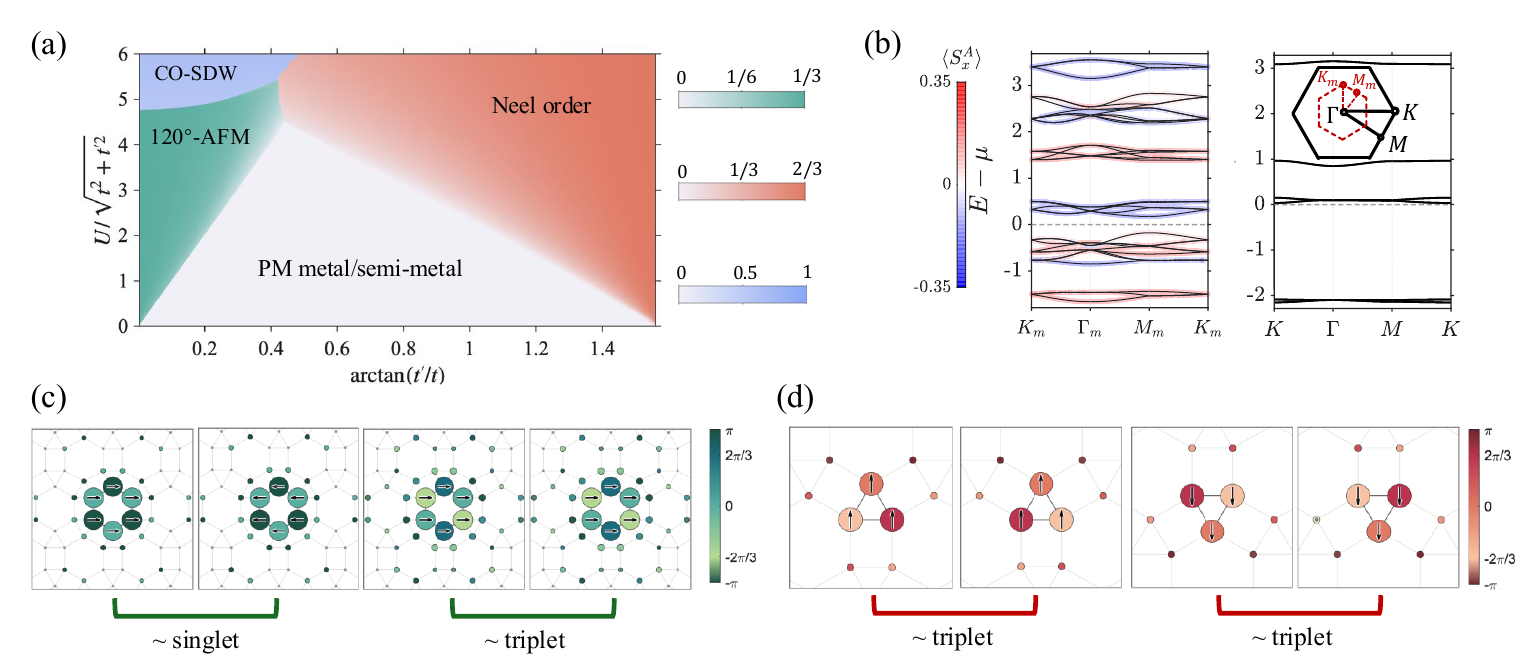} 
   \caption{Mean-field calculations of magnetism in the ruby-lattice Hubbard model at $n_e=4$. 
    (a) Mean field phase diagram with the $120^\circ$-AFM (green), N\'eel (red), CO-SDW (blue) and PM (gray) phases. The paramagnetic phase is metallic or semimetallic. The colorbars show magnitudes of magnetic moments $2|\langle\hat{\boldsymbol{s}}_i\rangle|$ on atomic sites.   
    (b) Mean-field band structures for the insulating $120^\circ$-AFM (left) and N\'eel order (right) calculated at $(t,t',U)=(1,0.2,4), (0.2,1,5)$ respectively. The color in the $120^\circ$-AFM spectrum represents $\langle \hat{S}_x\rangle$ of each Bloch state on a hexamer in the magnetic supercell. The dashed horizontal line marks the Fermi level. The inset shows the high-symmetry path in BZ (black) and magnetic BZ (red). (c)(d) Wannier orbitals in $120^\circ$ phase (c) and N\'eel phase (d) constructed from mean field bands in (b). Area of each circle is proportional to $\sqrt{W_n^{\dagger}W_n}$ and colors show the phases of the spinor components along the directions of corresponding moments. Arrows indicate the directions of the local spin contribution $W_n^\dagger\boldsymbol{\sigma}W_n/2$. For each cluster, the two electrons on the active Wannier orbitals form an approximate triplet retaining the molecular characters. Four Wannier orbitals of a hexamer in a magnetic supercell are shown, and others can be found in Supplemental Material~\cite{SupplementalMaterial}, exhibiting $120^\circ$ order.}
   \label{fig:ruby}
\end{figure*} 

\section{Molecular-orbital magnetism on the ruby lattice}
\label{sec:ruby}
We now turn to the ruby lattice at a fixed electron filling on which different molecular units yield non-bipartite or bipartite emergent lattices. Consequently, AFM exchange between molecular moments results in frustrated and unfrustrated magnetism respectively.
\par As illustrated in Fig.\ref{fig:ruby schemes}(a), we consider a tight-binding Hubbard model on the ruby lattice with intertriangle hopping $t$, intratriangle hopping $t'$ and onsite Hubbard interaction $U$. The Hamiltonian for our model reads
\begin{equation}
\begin{aligned}
\hat{H}=\sum_{\langle i,j\rangle,\sigma} (t_{ij}\hat{c}_{i\sigma}^{\dagger}\hat{c}_{j\sigma}+\text{h.c.})+U\sum_{\boldsymbol{R}\alpha}\hat{n}_{\boldsymbol{R}\alpha\uparrow}\hat{n}_{\boldsymbol{R}\alpha\downarrow},
\end{aligned}
\label{ruby Hamiltonian}
\end{equation}
where $t_{ij}=t$ if $\langle i,j\rangle$ belongs to inter-triangle bonds and $t_{ij}=t'$ for intra-triangle bonds. Let $\boldsymbol{R}$ label unit cell and $\alpha$ the atomic site within it.  Within a hexamer, the sites are labeled by $\alpha\in\{1,\ldots,6\}$; within trimers 1 and 2, they are labeled by $\alpha\in\{a_1,b_1,c_1,a_2,b_2,c_2\}$. We fix the filling at four electrons per six-site unit cell ($n_e=4$).
 When $t\gg t'$, the lattice decomposes into weakly coupled hexamers centered at $1a$ (green shaded clusters in Fig.\ref{fig:ruby schemes}(a)) each of which contains four electrons. MOs of these hexagons form an emergent triangular lattice which is non-bipartite. For $t'\gg t$, trimers centered at 2b  filled with two electrons are the natural clusters (red shaded clusters in Fig.\ref{fig:ruby schemes}(a)) and a bipartite honeycomb lattice arises from these trimers. Following the same separation of energy scales in Sec.\ref{sec:warmup}, we consider $t\gg U\gg t'$ for weakly coupled hexamers and $t'\gg U\gg t$ for weakly coupled trimers (1a and 2b limits in Fig.\ref{fig:ruby schemes}(b), respectively) under $n_e=4$. 
\par \textit{1a limit}. We consider the limit $t\gg U$ first. Molecular orbitals on a single hexagon are 
\begin{equation}
\begin{aligned}
\ket{\phi_\mu}=\dfrac{1}{\sqrt{6}}\sum_{\alpha=1}^6 e^{ik_\mu(\alpha-1)}\ket{\alpha},
\end{aligned}
\end{equation}
with $\mu=0,...,5$ and $k_\mu=\mu\pi/3$. MO levels are $E_\mu=2t\cos k_\mu$ with the lowest energy $E_3=-2t$. 
At $U=0$, the four electrons form a low-energy manifold with $\ket{\phi_3}$ doubly occupied and two electrons distributed over the degenerate $\phi_2,\phi_4$ orbitals, comprising one triplet and three singlets. For $U\ll t$, we project Hubbard interactions $\hat{H}_U$ onto these occupied orbitals and neglect the pair-hopping terms between $\phi_3$ and $\phi_2,\phi_4$ ($\hat{H}_P\sim \frac{U}{6}\phi_{2\sigma}^{\dagger}\phi_{4\bar{\sigma}}^{\dagger}\phi_{3\bar{\sigma}}\phi_{3\sigma}+\text{h.c.}$), since they involve excitations of energy $\mathcal{O}(t)$ and contribute only $\mathcal{O}(U^2/t)$ corrections to low-energy Hamiltonian. Equivalently, we restrict projected Hamiltonian to subspace in which $\phi_3$ is doubly occupied ($n_{3\uparrow}=n_{3\downarrow}=1$), yielding the effective interaction
\begin{equation}
\begin{aligned}
\hat{H}_{U}^{1a}=-\dfrac{U}{6}\sum_{\mu=2,3,4}\hat{n}_{\mu\uparrow}\hat{n}_{\mu\downarrow}-\dfrac{U}{6}\hat{\boldsymbol{S}}^2_{\text{tot}}+\dfrac{U}{24}\hat{N}^2+\dfrac{U}{12}\hat{N},
\end{aligned}
\label{Hu 1a}
\end{equation}
where $\hat{n}_{\mu\sigma}=\phi_{\mu\sigma}^{\dagger}\phi_{\mu\sigma}$ is the particle-number operator on $\ket{\phi_{\mu}\sigma}$, $\hat{\boldsymbol{S}}_{\text{tot}}$ is the total spin operator and $\hat{N}=\sum_{\mu\sigma}\hat{n}_{\mu\sigma}$ is the total particle number operator in a hexagonal cluster. The effective Hund coupling $-(U/6)\hat{\boldsymbol{S}}_{\mathrm{tot}}^2$ selects the spin-triplet states as the ground-state manifold. The resulting spin-$1$ hexamer moment is carried by the two electrons in the active MOs $\phi_2,\phi_4$. After turning on $t'$, second-order perturbation theory for neighboring hexamer triplets yields the effective spin interaction 
\begin{equation}
\begin{aligned}
\hat{H}_{1a}=J_{1a}\sum_{\langle i,j\rangle}(\hat{\boldsymbol{S}}_i\cdot \hat{\boldsymbol{S}}_j-1)+\text{const.},
\end{aligned}
\label{eq: ruby 1a Hamiltonian}
\end{equation}
and $\langle i,j\rangle$ here denotes a nearest-neighbor bond on the emergent triangular lattice from hexamer MOs. Most importantly, $J_{1a}=5t'^2/6U-t'^2/9t+2t'^2/(18t+9U)\xrightarrow{U/t\ll 1}5t'^2/6U\textgreater 0$ results in an antiferromagnetic spin-$1$ Heisenberg model on the triangular lattice [Fig.\ref{fig:ruby schemes}(b)] where the moments cannot be antiparallel on all three bonds of a triangle and are frustrated. The resulting frustrated spin-$1$ Heisenberg model supports $120^\circ$ order~\cite{Gotze2016TriangularHeisenberg}.
\par \textit{2b limit}. The same procedure is applied in the $2b$ limit with two electrons per trimer. The three MOs form two degenerate $E$ orbitals with energy $-t'$ and a nondegenerate $A_1$ orbital with energy $2t'$ under $D_3$ symmetry. For $t'>0$ and $t'\gg U$, we project out the higher-energy $A_1$ orbital and obtain the Hubbard interaction within $E$ subspace:
\begin{equation}
\begin{aligned}
\hat{H}_U^{2b}=-\dfrac{U}{3}\sum_{\mu=\pm}\hat{n}_{\mu\uparrow}\hat{n}_{\mu\downarrow}-\dfrac{U}{3}\hat{\boldsymbol{S}}_{\text{tot}}^2+\dfrac{U}{12}\hat{N}^2+\dfrac{U}{6}\hat{N},
\end{aligned}
\label{Hu 2b}
\end{equation}
where $\mu\in\{E_+,E_-\}$ labels two $E$ orbitals. As in the 1a limit, the effective Hund's coupling is $-U/3\hat{\boldsymbol{S}}_{\text{tot}}^2$ and prefers triplet ground states. Thus the two $E$ orbitals are each singly occupied, and two electrons form an effective spin-$1$ moment centered on the trimer. A small $t$ could induce the antiferromagnetic coupling between these spins, 
\begin{equation}
\begin{aligned}
\hat{H}_{2b}=J_{2b}\sum_{\langle i,j\rangle }(\hat{\boldsymbol{S}}_i\cdot \hat{\boldsymbol{S}}_j-1),\ \  J_{2b}=\dfrac{5t^2}{3U}\textgreater 0.
\end{aligned}
\label{eq: 2b Hamiltonian}
\end{equation}
\par As we illustrated in Fig.\ref{fig:ruby schemes}(b), the positive exchange $J_{2b}$ yields a spin-$1$ antiferromagnet on the bipartite honeycomb lattice, which supports N\'eel order~\cite{Gong2015HoneycombSpinOne}, in contrast to the frustrated magnetism in the 1a hexamer limit. Taken together, these results demonstrate that with the same atomic lattice and fixed electron filling, varying $t'/t$ can switch between frustrated and unfrustrated antiferromagnetism through changing the molecular units and the corresponding MO lattices.
\par We now explore how this molecular-orbital picture extends beyond the two controlled limits. For isolated hexamers and trimers, exact diagonalization (ED) finds triplet ground states separated from excited states with finite gaps for finite positive $U$~\cite{SupplementalMaterial}. The two clusters nevertheless respond differently to increasing $U$: the triplet in a trimer formed by two $E$ MOs remains an exact ground state for all $U>0$, whereas the hexamer ground state acquires increasing admixtures of other MO configurations, reducing the weight of the triplet MO   configuration obtained in the 1a limit while retaining total spin $S=1$. Moreover, finite intercluster hopping hybridizes the compact MOs and broadens their discrete levels into dispersive energy bands. The MOs can then evolve into localized Wannier orbitals with tails extending onto neighboring clusters. Intercluster hopping also induces charge fluctuations between clusters that can weaken molecular moments. ED of two coupled hexamers or trimers finds nondegenerate singlet ground states for positive $U$ and nonzero intercluster hopping, while increasing $U$ suppresses these charge fluctuations and stabilizes substantial local moments. These cluster results motivate an investigation of magnetism and its molecular-orbital character on the full ruby lattice over a broader range of $U$ and $t'/t$.
\par \textit{Mean field analysis}. We therefore turn to a spin-channel mean field analysis of the original Hubbard model to explore its magnetic phase diagram at finite hopping ratio $t'/t$ and a broader range of $U$. 
 Hubbard interaction $\hat{H}_U$ in Eq.\eqref{ruby Hamiltonian} can be rewritten as  
\begin{equation}
\begin{aligned}
\hat{H}_U=-\dfrac{2}{3}U\sum_{\boldsymbol{R}\alpha}\hat{\boldsymbol{s}}_{\boldsymbol{R}\alpha}\cdot \hat{\boldsymbol{s}}_{\boldsymbol{R}\alpha}+\dfrac{U}{2}\hat{N},
\end{aligned}
\end{equation} 
where $\hat{\boldsymbol{s}}_{\boldsymbol{R}\alpha}$ is the spin operator for site $\alpha$ in unit cell $\boldsymbol{R}$. Mean field approximation sets $\hat{\boldsymbol{s}}_{\boldsymbol{R}\alpha}\cdot \hat{\boldsymbol{s}}_{\boldsymbol{R}\alpha}\simeq 2\langle \hat{\boldsymbol{s}}_{\boldsymbol{R}\alpha}\rangle \cdot \hat{\boldsymbol{s}}_{\boldsymbol{R}\alpha}-|\langle \hat{\boldsymbol{s}}_{\boldsymbol{R}\alpha}\rangle |^2$ and yields a quadratic mean field Hamiltonian which is easy to diagonalize. We calculate charge density $\langle \hat{n}_{\boldsymbol{R}\alpha}\rangle$ and local spin $\langle \hat{\boldsymbol{s}}_{\boldsymbol{R}\alpha}\rangle$ from bands of mean field Hamiltonian and iterate to self-consistency~\cite{SupplementalMaterial,Inui1991HartreeFock,Scholle2023MeanField}. By examining several magnetic ansatze including $120^\circ$ AFM and N\'eel order and multiple initial values of the order parameters, we compare the energies of the converged solutions and use the lowest-energy state at each parameter point to construct the mean-field phase diagram in Fig.~\ref{fig:ruby}(a). 
\par We find the predicted $120^\circ$-AFM and N\'eel order in the green and red regions of phase diagram. In both phases, the local moments have equal magnitudes on the atomic sites, which is compatible with the distribution of corresponding molecular orbitals. The phase boundary between $120^\circ$-AFM and Neel order is about $t'/t\sim 0.45$ ($\arctan(t'/t)\sim 0.43$). At $t'/t$ away from 1a and 2b limits, a finite threshold $U_c$ is needed to support finite magnetic moments due to stronger charge fluctuations in this region. For $U\leq U_c$ we get the paramagnetic phase, which is metallic or semi-metallic depending on the band structure around Fermi level. All N\'eel and most of $120^\circ$-AFM mean-field solutions are insulating, with a single-particle gap at the Fermi level[Fig.\ref{fig:ruby}(b)]. Near the boundary with the Néel phase, however, the gap in $120^\circ$-AFM phase closes. At larger $U$, a charge-order spin-density wave (CO-SDW) state appears in the blue region in Fig.\ref{fig:ruby}(a), briefly discussed in Supplemental Material\cite{SupplementalMaterial}. In this state, electrons are distributed nonuniformly over the sites in a hexamer, with moments forming a collinear AFM pattern. This may reflect limitations of mean field theory and the stability of this phase requires more precise many-body calculations. 
\par We next focus on insulating states in the green and red regions in Fig.\ref{fig:ruby}(a) to show that features of molecular moments persist, with compact MOs evolving into localized Wannier orbitals at finite intercluster coupling. At representative parameters shown in Fig.\ref{fig:ruby}(c)(d), we construct the localized Wannier spinors from the occupied mean field bands using a projection method with symmetric L\"owdin orthonormalization~\cite{Solovyev2007Wannier}, starting from compact MO spinors as trial orbitals. For each spinor $W_n(i)$, $i=(\boldsymbol{R},\alpha)$, its contribution to the spin density is $\langle \hat{\boldsymbol{s}}_i\rangle_n =W_n^{\dagger}(i)\boldsymbol{\sigma}W_n(i)/2$. As shown in Fig.\ref{fig:ruby}(c)(d), we find that for each cluster, the two electrons occupying two active Wannier orbitals with parallel spins form a nearly pure triplet. And in hexamers, the two additional occupied orbitals have nearly cancelling spin contributions. These triplet states formed by magnetic Wannier orbitals provide the counterparts of compact-MO moments at finite couplings and exhibit the predicted $120^\circ$ and N\'eel orders. Their spin densities are primarily localized on the parent clusters while their tails extending onto neighboring clusters permit charge and spin fluctuations and can reduce the moments within original clusters.
       
\section{Low-energy magnetic excitations and molecular fingerprints}
\label{sec:formfactor}
\begin{figure}[tb]
   \centering
   \includegraphics[width=\columnwidth]{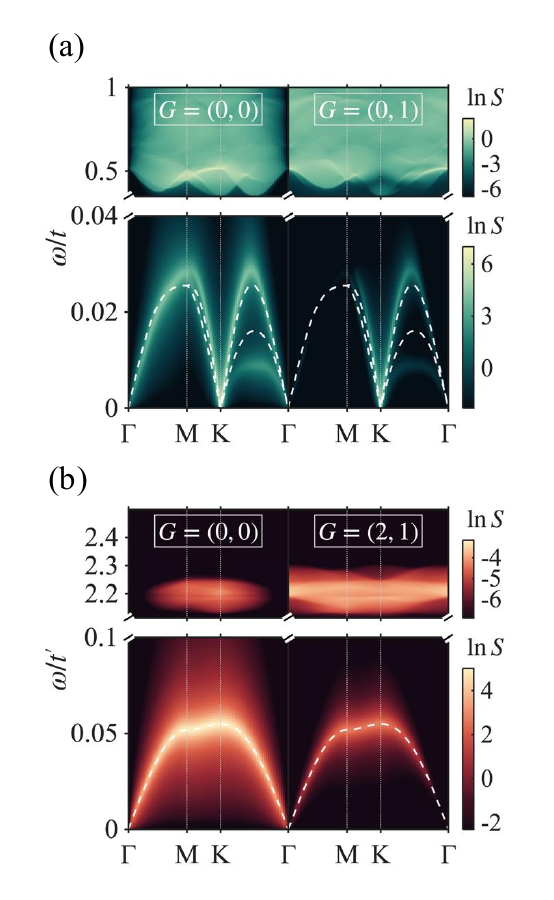}
   \caption{RPA spectra $\ln\mathcal{S}(\boldsymbol{q}+\boldsymbol{G},\omega)$ along high-symmetry path for (a) the $120^\circ$-AFM phase at $(t,t',U)=(1,0.2,4)$ and (b) the N\'eel phase at $(t,t',U)=(0.2,1,5)$. BZs are labeled by $(m,n)$, where $\boldsymbol{G}=m\boldsymbol{b}_1+n\boldsymbol{b}_2$. 
   The frequency axes are broken, with separate color scales for the low and high-energy windows.
   White curves show linear spin-wave dispersions of the spin-$1$ Heisenberg models on the corresponding emergent lattices, with one exchange constant fitted to the RPA dispersion for each phase. The low-energy modes have the same dispersions but different spectral weights in the two BZs. At higher frequencies, the DSF exhibits the continuum from interband particle-hole excitations.}
   \label{Bragg Peaks}  
\end{figure} 
\par We next identify the spectroscopic fingerprints of molecular-orbital magnetism in low-energy dynamic structure factor (DSF), which is probed by inelastic neutron scattering~\cite{Squires2012NeutronScattering}. The DSF describes collective fluctuations of moments carried by MOs on emergent lattices through correlations of the microscopic spin density, and therefore contains information about both the dynamics of molecular moments and their spatial profiles.
\par Our starting point is the zero-temperature DSF,
\begin{equation}
\begin{aligned}
\mathcal{S}^{\mu\nu}(\boldsymbol{q},\omega)\equiv \dfrac{1}{N_c}\langle \hat{S}^{\mu}(\boldsymbol{q})\delta(\omega-\hat{H}+E_0)\hat{S}^{\nu}(-\boldsymbol{q})\rangle,
\end{aligned}
\label{eq:DSF definition}
\end{equation}
where $N_c$ is the number of unit cells, $\hat{S}^\mu(\boldsymbol{q})$ is the Fourier component of the microscopic spin density and the expectation value is taken in the ground state of $\hat{H}$ with energy $E_0$. The elastic contribution at $\omega=0$ gives magnetic Bragg peaks. In the ruby-lattice model, the peaks occur at $K/K'$ points and $\Gamma$ points for $120^\circ$ order on the emergent triangular lattice and N\'eel order on the emergent honeycomb lattice respectively. The central $\Gamma$ peak is absent as the opposite moments on trimers cancel within a unit cell.

\par To examine excitations of these molecular spins, we calculate the DSF in $\mathcal{S}=\mathcal{S}^{xx}+\mathcal{S}^{yy}$ channel using RPA based on mean-field magnetic states~\cite{Schrieffer1989SpinFluctuations,Willsher2023MagneticExcitations}. The resulting spectra show coherent spin waves at low energies and a continuum of interband particle-hole excitations at higher energies [Fig.~\ref{Bragg Peaks}(a,b)]. We focus on the low-energy gapless Goldstone modes emanating from magnetic Bragg peaks below the particle-hole continuum. Fig.\ref{Bragg Peaks} compares the RPA calculated low-energy $\mathcal{S}_{\text{RPA}}(\boldsymbol{q},\omega)$ with linear spin-wave dispersions of nearest-neighbor AFM spin-1 Heisenberg models on the emergent triangular and honeycomb lattices respectively. For each phase, we fit a single Heisenberg spin coupling $J$ to the RPA excitation energies. The agreement supports a low-energy picture in which the excitations do not disrupt the effective molecular spins but instead involve their collective fluctuations on the emergent lattices.
\par The low-energy DSF in Fig.\ref{Bragg Peaks} shows the same spin-wave dispersions in different BZs, but the corresponding spectral weights vary substantially. We now demonstrate that such intensity modulation encodes the information about spatial spin density of molecular moments, specifically the Fourier transform of their spin density, providing the spectroscopic clues to molecular-orbital magnetism. 
\par To analyze these molecular fingerprints in low-energy DSF more generally, we consider a lattice of weakly coupled clusters with Hamiltonian
\begin{equation}
\begin{aligned}
\hat{H}=\sum_{\boldsymbol{R}\lambda}\hat{H}_{\boldsymbol{R}\lambda}+g\hat{H}',
\end{aligned}
\label{general Hamiltonian}
\end{equation}
where $\hat{H}_{\boldsymbol{R}\lambda}$ is the Hamiltonian of $\lambda$-cluster in unit cell $\boldsymbol{R}$ (e.g., hexamer or trimer in the ruby-lattice model) and $g\hat{H}'$ is weak coupling between these clusters.  We assume that ground-state manifold $V_{\boldsymbol{R}\lambda}$ of $\hat{H}_{\boldsymbol{R}\lambda}$ is a single isolated spin-$T_\lambda$ multiplet with no additional orbital degeneracy. Excitations out of these cluster ground-state multiplets cost at least an energy  $\Delta$ in the decoupled limit.  This gap, which we call the molecular spin stiffness, protects the integrity of the cluster moments at low energies as a collective spin degree of freedom. We require $g\ll\Delta$ to maintain a well-defined cluster spin $T_\lambda$ on the emergent lattice formed by these clusters, as we have seen in Secs.~\ref{sec:warmup} and \ref{sec:ruby}. 
\par Let $P_{\boldsymbol{R}\lambda}$ project onto $V_{\boldsymbol{R}\lambda}$ of each cluster. The full low-energy subspace $\mathcal{V}_0=\bigotimes_{\boldsymbol{R}\lambda}V_{\boldsymbol{R}\lambda}$ then has projector $P_0=\prod_{\boldsymbol{R}\lambda}P_{\boldsymbol{R}\lambda}$. To describe low-energy physics in terms of these cluster spins, we apply the Schrieffer-Wolff (SW) transformation~\cite{Schrieffer1966Transformation,Bravyi2011SchriefferWolff} to construct the effective Hamiltonian $\hat{H}_{\text{eff}}=P_0\widetilde{\hat{H}}P_0$ and effective spin-density operator $\hat{S}^{\mu}_{\text{eff}}(\boldsymbol{q})=P_0\widetilde{\hat{S}}^{\mu}(\boldsymbol{q})P_0$ acting within $\mathcal{V}_0$. Together, they determine the low-energy DSF. The tildes denote SW transformed operators. Details are given in the Supplemental Material~\cite{SupplementalMaterial}. 
\par \textit{Spatial profiles and molecular form factors}.  
The low-energy projection of spin-density operator identifies the spatial profiles carried by each effective moment. For an isolated cluster, we suppress the unit-cell label $\boldsymbol{R}$ and denote its local spin-density operator by $\hat{s}_{\lambda}^\mu(\boldsymbol{u})$. Within the ground-state multiplet $V_\lambda$, the Wigner-Eckart theorem gives~\cite{SupplementalMaterial,Waldmann2005EffectiveOperators}  
\begin{equation}
P_\lambda\hat{s}_\lambda^\mu(\boldsymbol{u})P_\lambda
=\rho_\lambda(\boldsymbol{u})\hat{T}_\lambda^\mu,
\label{WE projection}
\end{equation}
where $\hat{T}_\lambda^\mu$ is the effective molecular-spin operator. This identity separates the spatial profile $\rho_{\lambda}(\boldsymbol{u})$ from the effective spin $\hat{T}_{\lambda}^{\mu}$. The coefficient $\rho_{\lambda}$ can be obtained from the spin density of the maximally polarized state in multiplet as 
$\rho_\lambda(\boldsymbol{u})
=\bra{T_\lambda,T_\lambda}
\hat{s}_\lambda^z(\boldsymbol{u})
\ket{T_\lambda,T_\lambda}
/{T_\lambda}$.
Its normalization, $\int d\boldsymbol{u}\rho_{\lambda}(\boldsymbol{u})=1$, ensures that integrated projected spin density reproduces the total molecular spin. The spatial spin density $\rho_{\lambda}$ contributes to momentum dependence of low-energy DSF through its Fourier transform, the molecular form factor,
\begin{equation}
F_\lambda(\boldsymbol{q})
=\int d\boldsymbol{u}\,
e^{-i\boldsymbol{q}\cdot\boldsymbol{u}}
\rho_\lambda(\boldsymbol{u}),
\end{equation}
In the discrete lattice description, $\rho_{\lambda}$ is specified by coefficients $\Gamma_{\lambda\alpha}$, the normalized
spin-density weights at site $\alpha$, from $P_\lambda\hat{s}^{\mu}_{\lambda\alpha}P_{\lambda}=\Gamma_{\lambda\alpha}\hat{T}^{\mu}_{\lambda}$. The corresponding form factor is $F_\lambda(\boldsymbol{q})
=\sum_\alpha\Gamma_{\lambda\alpha}
e^{-i\boldsymbol{q}\cdot\boldsymbol{d}_{\lambda\alpha}}$ where $\boldsymbol{d}_{\lambda\alpha}$ is the position of site $\alpha$ relative to the cluster center. 
\par  To leading order in the intercluster coupling, the effective spin density on the lattice is
\begin{equation}
\begin{aligned}
\hat{S}^{\mu}_{\text{eff}}(\boldsymbol{q})=\sum_{\lambda}F_{\lambda}(\boldsymbol{q})e^{-i\boldsymbol{q}\cdot\boldsymbol{\rho}_{\lambda}}\hat{T}_{\lambda}^{\mu}(\boldsymbol{q})
\end{aligned}
\end{equation}
where $\boldsymbol{\rho}_\lambda$ is the center
of cluster-$\lambda$ in a unit cell.
The dynamical correlations of effective spin density yield the low-energy DSF at $\omega<\Delta$:

\begin{figure*}[t]
  \centering
   \includegraphics[width=\textwidth]{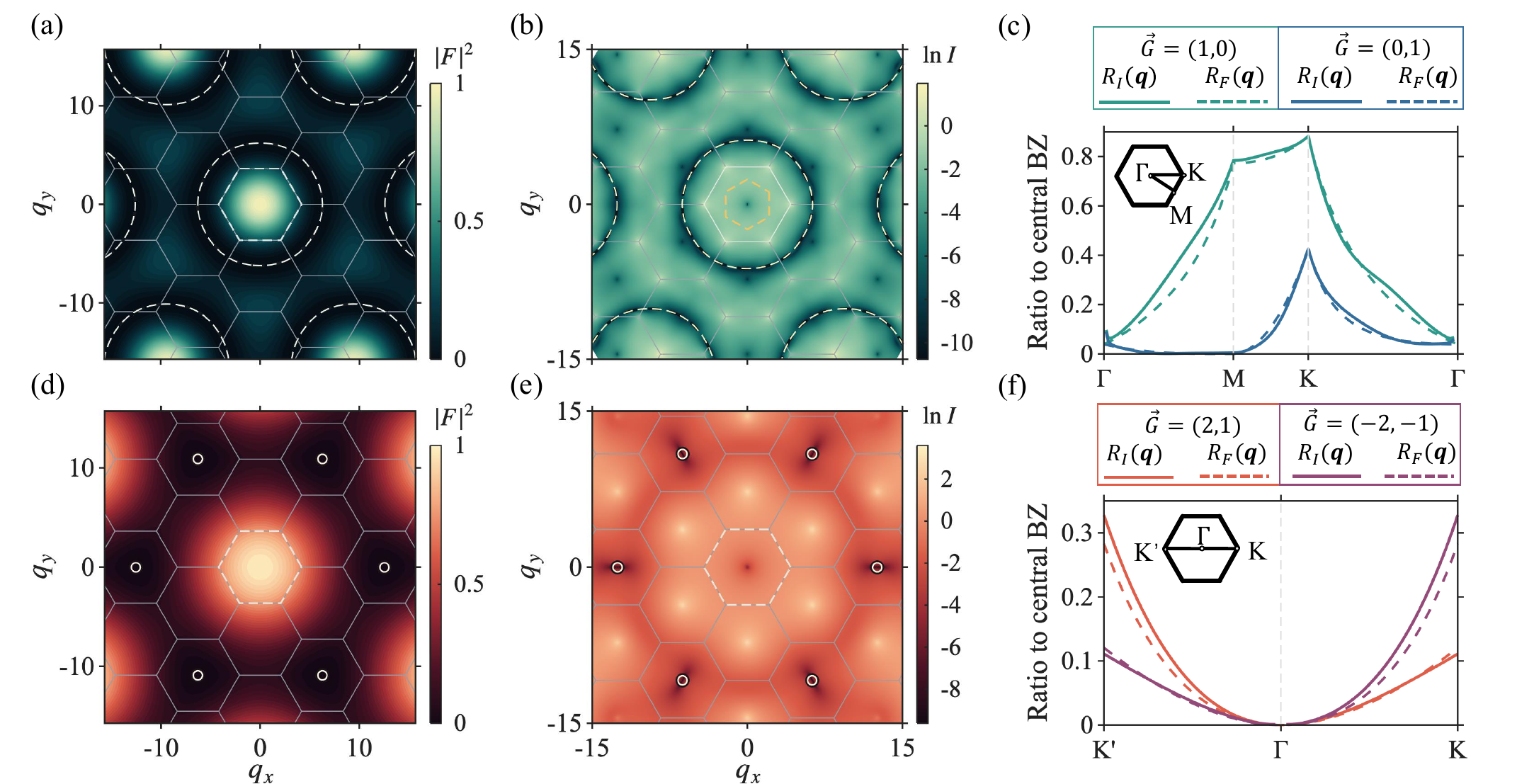}
   \caption{Molecular form-factor fingerprints of molecular-orbital magnetism on emergent lattices.
   Panels (a)--(c) correspond to the $120^\circ$-AFM phase of hexamer moments on the triangular MO lattice at $(t,t',U)=(1,0.2,4)$; panels (d)--(f) correspond to the N\'eel phase of trimer moments on the honeycomb MO lattice at $(t,t',U)=(0.2,1,5)$.
   (a),(d) Squared molecular form factors $|F(\boldsymbol{q})|^2$, normalized to $|F(0)|^2=1$.
   (b),(e) Logarithm of $I_{\text{RPA}}(\boldsymbol{q})$ in the same spin channel, using the low-energy frequency windows of lowest spin-wave branches and excluding the elastic contribution.
   White dashed contours in (a),(b) and white circles in (d),(e) mark the zeros of the molecular form factors. The RPA intensity in (b),(e) exhibits minima near these zeros. Gray lines outline the BZs.
   (c),(f) Comparison of $R_I^{\boldsymbol{G}}(\boldsymbol{q})$ (solid) and $R_F^{\boldsymbol{G}}(\boldsymbol{q})$ (dashed) along the paths shown in the insets.
   Reciprocal vectors are labeled by $(m,n)$ with $\boldsymbol{G}=m\boldsymbol{b}_1+n\boldsymbol{b}_2$. }
   \label{fig:formfactor}
\end{figure*}

\begin{equation}  
\begin{aligned}
\mathcal{S}^{\mu\nu}_{\text{low}}(\boldsymbol{q},\omega<\Delta)=\sum_{\lambda\lambda'}e^{-i\boldsymbol{q}\cdot (\boldsymbol{\rho}_{\lambda}-\boldsymbol{\rho}_{\lambda'})}F_{\lambda}(\boldsymbol{q})F_{\lambda'}^*(\boldsymbol{q})\mathcal{C}^{\mu\nu}_{\lambda\lambda'}(\boldsymbol{q},\omega),
\end{aligned}
\label{eq: low DSF}
\end{equation}
where $\mathcal{C}_{\lambda\lambda'}^{\mu\nu}(\boldsymbol{q},\omega)\equiv \langle \hat{T}^{\mu}_{\lambda}(\boldsymbol{q})\delta(\omega-\hat{H}_{\text{eff}}+\mathcal{E}_0)\hat{T}^{\nu}_{\lambda'}(-\boldsymbol{q})\rangle/N_c$ is the dynamic correlation of $\lambda$ and $\lambda'$ cluster spins, evaluated in the ground state of $\hat{H}_{\text{eff}}$ with energy $\mathcal{E}_0$. The correlations $\mathcal{C}^{\mu\nu}_{\lambda\lambda'}$, governed by $\hat{H}_{\mathrm{eff}}$, contain the dynamics of molecular moments on the emergent lattices (e.g. spin waves in Fig.\ref{Bragg Peaks}). The form factors $F_\lambda(\boldsymbol{q})$ encode how their spatial profiles modulate the momentum dependence of the corresponding spectral weight. 
\par We then identify two spectroscopic fingerprints of molecular spin-density profiles that provide direct experimental clues to MO magnetism. First, a pronounced suppression of the low-energy DSF occurs at momenta where all molecular form factors vanish. At these momenta, the leading order of $\hat{S}_{\text{eff}}^{\mu}(\boldsymbol{q})$ is then zero, resulting in exact extinction of low-energy DSF at this order irrespective of the effective spin correlations. We denote this common zero set of molecular form factors by $Z_F$. The locations of these extinctions are therefore fixed by the internal spatial spin-density distribution of clusters. Corrections from weak intercluster couplings may lift the exact zeros while leaving pronounced intensity minima around $Z_F$. Such persistent suppression can serve as intrinsic spectroscopic fingerprints and implies MO magnetism.
\par Second, for one cluster per unit cell, a further quantitative test follows from the factorized DSF. We can omit the $\lambda$ index to write
\begin{equation}
\begin{aligned}
\mathcal{S}_{\text{low}}(\boldsymbol{q},\omega)=|F(\boldsymbol{q})|^2\mathcal{C}(\boldsymbol{q},\omega).
\end{aligned}
\end{equation}
The spin correlation $\mathcal{C}(\boldsymbol{q},\omega)$ is invariant under $\boldsymbol{q}\rightarrow \boldsymbol{q}+\boldsymbol{G}$ where $\boldsymbol{G}$ is a reciprocal lattice vector. Therefore the effective correlation $\mathcal{C}(\boldsymbol{q},\omega)$ cancels in the ratio of the DSFs at $\boldsymbol{q}$ and $\boldsymbol{q}+\boldsymbol{G}$. This ratio is determined by molecular form factor and is therefore independent of frequency throughout the low-energy range. Such a frequency-independent ratio can itself provide another spectroscopic clue to MO magnetism. 
\par To compare spectral weights, particularly for spectra containing sharp coherent modes, we integrate
the DSF over a fixed low-energy window to collect their
spectral weights. The integrated intensity is 
\begin{equation}
\begin{aligned}
I(\boldsymbol{q})\equiv \int_{0^+}^{\Omega}\mathcal{S}_{\text{low}}(\boldsymbol{q},\omega)d\omega
\end{aligned}
\end{equation}
where $\Omega$ is a frequency cutoff to exclude high-energy excitations and $0^+$ excludes elastic contribution. The deep minima of $I(\boldsymbol{q})$ around $Z_F$ support the MO description of magnetism. Moreover, we define two ratios as 
\begin{equation}
\begin{aligned}
& R_I^{\boldsymbol{G}}(\boldsymbol{q})\equiv\dfrac{I(\boldsymbol{q}+\boldsymbol{G})}{I(\boldsymbol{q})},\\
& R_F^{\boldsymbol{G}}(\boldsymbol{q})\equiv \dfrac{|F(\boldsymbol{q}+\boldsymbol{G})|^2}{|F(\boldsymbol{q})|^2},\ \boldsymbol{q}\in \text{1st BZ}.
\end{aligned}
\label{ratios in BZ}
\end{equation}
So factorized DSF gives \(R_I \simeq R_F\). These fingerprints in low-energy DSF can inspire one to come up with an MO description of the magnetism in a material. A proposed molecular profile can then be experimentally tested by comparing its predicted $Z_F$ with intensity minima and $R_F$ with the ratio extracted from the spectrum. For multiple clusters per unit cell, interference between clusters in a unit cell generally makes the intensity ratio $R_I^{\boldsymbol{G}}(\boldsymbol{q})$ depend on their spin correlations.   
\par We demonstrate these predictions above in the ruby-lattice model of Sec.\ref{sec:ruby} with lattice constant $a=1$, side length of hexagon $l=a/(1.5\sqrt{3})$ and lattice vectors $\boldsymbol{a}_1=a(1,0)$, $\boldsymbol{a}_2=a(1/2,\sqrt{3}/2)$. We integrate $\mathcal{S}_{\text{RPA}}$ over a frequency window of spin-wave branches with lowest energy to obtain the integrated intensity $I_{\text{RPA}}(\boldsymbol{q})$. Fig.~\ref{fig:formfactor} shows the resulting $I_{\text{RPA}}(\boldsymbol{q})$ and the molecular form factors, comparing intensity minima with zeros of form factors and ratios $R_I$ with $R_F$. The close agreement confirms that signatures remain visible in the $120^\circ$-AFM and Néel phases beyond controlled limits.
\par In the 1a $120^\circ$-AFM phase, the form factor of a hexagon cluster with spin-$1$ is 
\begin{equation}
\begin{aligned}
F_{\hexagon}(\boldsymbol{q})=\dfrac{1}{6}\sum_{\alpha\in \hexagon}e^{-i\boldsymbol{q}\cdot\boldsymbol{d}_{\lambda\alpha}}\in\mathbb{R},
\end{aligned}
\label{hexagon form factor}
\end{equation}
which is purely real due to inversion symmetry about the cluster center. The condition $F_{\hexagon}(q_x,q_y)=0$ therefore imposes a single real constraint, generally giving lines of zeros as plotted in Fig.\ref{fig:formfactor}(a). The RPA intensity $I_{\text{RPA}}(\boldsymbol{q})$ exhibits clear minima that closely follow these zero contours in $F_{\hexagon}(\boldsymbol{q})$ [Fig.~\ref{fig:formfactor}(b)]. Eq.\eqref{ratios in BZ} is also verified in Fig.\ref{fig:formfactor}(c), showing that $R_I$ from $I_{\text{RPA}}(\boldsymbol{q})$ closely follows the $R_F$. This agreement shows that low-energy DSF is modulated by molecular form factor, which strongly supports a description in terms of spatially extended molecular moments. 
\par In the 2b N\'eel phase, there are two triangular clusters in a unit cell with complex-conjugate form factors, 
\begin{equation}
\begin{aligned}
F_{\triangle}(\boldsymbol{q})=\dfrac{1}{3}\sum_{\alpha\in\triangle}e^{-i\boldsymbol{q}\cdot\boldsymbol{d}_{\lambda\alpha}}=F_{\triangledown}^*(\boldsymbol{q}).
\end{aligned}
\end{equation}
$F_{\triangle}$ is generally complex so its zeros require both $\text{Re}(F_{\triangle})=0$ and $\text{Im}(F_{\triangle})=0$. These two real constraints in general yield isolated zeros as shown in Fig.\ref{fig:formfactor}(d). RPA calculated $\log I(\boldsymbol{q})$ has deep minima at these positions [Fig.~\ref{fig:formfactor}(e)], revealing the triangular spin-density profiles of the molecular moments. For the two-trimer unit cell, intersublattice interference prevents a direct application of the single-cluster ratio relation [Eq.\eqref{ratios in BZ}] at general $\boldsymbol{q}$. But we find a special path along $K'-\Gamma-K$ for reciprocal vectors $\boldsymbol{G}$ along the x axis where $R_I\simeq R_F$ holds. Along these paths, the two trimer form factors are equal and real at both $\boldsymbol{q}$ and $\boldsymbol{q}+\boldsymbol{G}$, while $e^{i\boldsymbol{G}\cdot(\boldsymbol{\rho}_{\triangle}-\boldsymbol{\rho}_{\triangledown})}=1$. Together with $\mathcal{C}_{\lambda\lambda'}(\boldsymbol{q}+\boldsymbol{G},\omega)=\mathcal{C}_{\lambda\lambda'}(\boldsymbol{q},\omega)$, we find only the common molecular form factor $F_{\triangle}(\boldsymbol{q})$ changes under $\boldsymbol{q}\rightarrow \boldsymbol{q}+\boldsymbol{G}$, yielding $\mathcal{S}_{\mathrm{low}}(\boldsymbol{q}+\boldsymbol{G},\omega)\simeq R_F^{\boldsymbol{G}}(\boldsymbol{q})\mathcal{S}_{\mathrm{low}}(\boldsymbol{q},\omega)$ and $R_I^{\boldsymbol{G}}(\boldsymbol{q})\simeq R_F^{\boldsymbol{G}}(\boldsymbol{q})$ along these paths [Fig.\ref{fig:formfactor}(f)].
  
\par We also notice that similar systems, where cluster spins come from strong intracluster exchange between individual spins at atomic sites in a cluster, could exhibit the same low-energy form factors and even the same collective modes as in molecular-orbital magnetism~\cite{Romhanyi2014EntangledTetrahedron,Luo2020LowEnergyMagnons}. Higher-energy electronic excitations can help distinguish these microscopic pictures. Molecular-orbital magnetism yields the higher-energy continuum because excitations can disrupt MOs and result in particle-hole excitations between bands, as shown in Fig.~\ref{Bragg Peaks}(a,b). These atomic spin models, however, lack the electronic degrees of freedom needed to describe this continuum. Therefore, molecular-orbital fingerprints, low-energy dynamics and the higher-energy particle-hole continuum together confirm a solid and reliable magnetic description using molecular spins on the emergent lattice.    
\par Corrections to the compact-MO description at finite intercluster coupling include the extension of molecular spin densities onto neighboring clusters, as captured by the Wannier orbitals discussed above. Additional RPA calculations in the Supplemental Material~\cite{SupplementalMaterial} nevertheless show that the molecular fingerprints persist beyond the perturbative limits, although the intensity minima become progressively less pronounced as intercluster hopping increases. Their visibility and agreement with the predictions from molecular form factors also indicate the accuracy of the MO description, as discussed in the Supplemental Material~\cite{SupplementalMaterial}. Together with the spin-wave comparison, the persistence of these fingerprints supports the conclusion that molecular moments remain meaningful magnetic degrees of freedom on the emergent lattices beyond the perturbative limits, and provides a spectroscopic test of MO magnetism in inelastic neutron scattering experiments.

\section{Conclusion and Discussion}
\label{sec:discussion}
We have demonstrated that MO formation provides a microscopic route to changing magnetic frustration on emergent lattices while the underlying atomic lattice and electron filling remain fixed. In the ruby-lattice Hubbard model, the onsite repulsion within active MOs favors spin-$1$ molecular moments in both the hexamer and trimer regimes. Varying $t'/t$ changes their emergent lattice from triangular to honeycomb, supporting frustrated $120^\circ$ and unfrustrated N\'eel antiferromagnetism respectively. 

\par Spatially extended spin density of these moments leaves detectable characteristic signatures in the low-energy DSF through the molecular form factors. Within the low-energy description, the common zeros of these form factors produce suppressed low-energy spectral weight independently of magnetic order or nature of the low-energy excitations. For one cluster per unit cell, or along selected momentum paths for two trimers per unit cell, spectral ratios are independent of frequency and the integrated intensity ratios satisfy $R_I\simeq R_F$. These relations provide a basis for identifying molecular-orbital magnetism in inelastic neutron scattering experiments.
     
\par The location of these moments at empty Wyckoff positions further suggests a connection to obstructed atomic insulators (OAIs). In an OAI, symmetric Wannier centers cannot be continuously moved onto atomic sites while preserving the band gap and protecting symmetries~\cite{Bradlyn2017TopologicalQuantumChemistry,Cano2018ElementaryBandRepresentations}. At partial filling, interactions in these obstructed atomic bands may stabilize local moments, giving rise to magnetism on a lattice formed by the Wannier centers. The recent identification of an obstructed atomic band crossing the Fermi level in $1H$-NbSe$_2$ motivates investigating this possibility in materials~\cite{Calugaru2026ObstructedAtomicBand}. The connection to OAIs also motivates exploring higher-order topology: whether analogous corner moments can arise from molecular orbitals, linking the organization of molecular moments to magnetic response at boundaries. 

\par Alongside these materials prospects, ultracold atoms offer a complementary platform for exploring MO magnetism. Ruby geometries have been realized in programmable Rydberg arrays~\cite{Semeghini2021ProbingSpinLiquids}, while advances in fermionic Hubbard simulators have enabled the study of collective magnetism at low temperatures~\cite{Xu2025CryogenicHubbard}. Bragg spectroscopy also provides access to dynamic spin correlations experimentally in ultracold Fermi gases~\cite{Senaratne2022SpinChargeSeparation}. These developments suggest a promising route toward realizing molecular-orbital magnetism and probing the associated collective excitations in fermionic ruby-lattice systems.

\begin{acknowledgments}
C.-R.W. thanks Ruiheng Liu for helpful and inspiring discussions. C.F. thanks Yuan Li for helpful discussions. This work was supported by the National Natural Science Foundation of China (NSFC) (Grant Nos. 12325404, 12547112, and 12188101), the National Key R\&D Program of China (Grant Nos. 2022YFA1403800 and 2023YFA1406704), and the Strategic Priority Research Program (B) of the Chinese Academy of Sciences (CAS) (Grant No. XDB1720000).
\end{acknowledgments}
 
\section*{Data availability}
The data that support the findings of this
article are not publicly available. The data are available from
the authors upon reasonable request.

\normalem 
\makeatletter
\def\@bibstyle{apsrev4-2-20authors} 
\makeatother
\bibliography{references}

\end{document}